**Disparity-in-Differences: Extracting Hierarchical Backbones of Weighted Directed Networks**

Hyunuk Kim

Department of Management and Entrepreneurship,
Martha and Spencer Love School of Business, Elon University, Elon, NC 27244, USA

**Abstract**

Networks are useful representations for complex systems. Especially, heterogeneous and asymmetrical relations commonly found in complex systems can be converted to weighted directed edges between nodes. The disparity filter (Serrano et al., 2009) has successfully extracted backbones, sets of important edges, from empirical networks but is not designed to incorporate node-node dependency that may encode hierarchical relations. This paper proposes an extended disparity filter named "disparity-in-differences" that assigns a synthetic relation between two nodes if one depends relatively more on the other where the extent of asymmetric dependence is measured by the disparity between a normalized edge weight difference and an expected edge weight difference. For evaluation, the proposed method is applied to a journal citation network, a U.S. airport network, the Enron email network, and a world trade network. Compared to the disparity filter, the proposed approach better captures hierarchical relations that align well with journal quality ratings, airport hub categories by size, levels of management, and a core-periphery structure of countries, respectively.

1. **Introduction**

In complex systems, relations between entities are often not equally significant and not unilateral. Varying importance and directionality can be represented with numerical values and links from sources to targets. Networks of which edge has a weight and a direction (i.e., weighted directed networks) are useful to examine these relations. As many empirical weighted directed networks have dense connections between nodes, important edges (i.e., backbone) have been extracted for intuitive understanding of complex systems.

Several existing backbone extraction methods introduce null models that leverage structural characteristics of weighted directed networks to calculate a statistical significance of each edge and collect highly significant edges as a backbone (Coscia & Neffke, 2017; Dianati, 2016; Marcaccioli & Livan, 2019; Serrano et al., 2009). Backbones extracted from these methods retain important connections better than naive approaches that remove edges by a threshold on edge weights. However, they do not explicitly consider asymmetric dependence that may be associated with hierarchical relations.

This paper proposes a new method named "disparity-in-differences" that measures a statistical significance of asymmetric dependence, which is defined as the difference of normalized edge weights between two nodes. The term "disparity" is adopted from the disparity filter (Serrano et al., 2009), a popular network backbone extraction method. The core idea of the proposed approach is simple and in line with established literature. For two nodes $i$ and $j$, the normalized weight of an edge $i \rightarrow j$ is considered as the extent that $i$ depends on $j$ and vice versa. According to the power-dependence relation concept (Emerson, 1962), the dependence of $i$ on $j$ is equal to the power of $j$ over $i$. In this framework, power imbalance between $i$ and $j$ can be represented as the difference in dependencies that was even used for quantifying relationships between industries (Casciaro & Piskorski, 2005). The difference between the normalized weights of $i \rightarrow j$ and $j \rightarrow i$ would thus serve as a proxy for hierarchical relation.

## 2. Methods and Materials

a. The Disparity Filter (Serrano et al., 2009)

$p_{ij}$, a normalized weight of an edge $i \rightarrow j$, is calculated as $w_{ij}/\sum_{n \in N_i} w_{in}$ where $w_{ij}$ is the edge weight of $i \rightarrow j$ and $N_i$ is the set of $i$'s neighbors. The null model of the disparity filter assumes that edge weights are uniformly distributed over neighbors. For a node with $k$ neighbors, normalized edge weights are sampled from a Dirichlet distribution, $Dir(\alpha_1 = 1, \alpha_2 = 1, \dots, \alpha_k = 1)$. A normalized weight then follows a Beta distribution, $Beta(1, k-1)$, where the probability density function $f(x)$ is $(k-1)(1-x)^{k-2}$. By integrating the probability density function from 0 to $p_{ij}$, we can obtain $\alpha_{ij}$, the statistical significance of $p_{ij}$ based on the null model. If $p_{ij}$ is significantly higher than expectations, $i \rightarrow j$ is included in a backbone. Throughout the paper, a threshold of 0.01 was used so that edges of which $\alpha$ is lower than or equal to 0.01 were extracted as a backbone.

b. Disparity-in-Differences

For two nodes $i$ and $j$, we can calculate $p_{ij}$ and $p_{ji}$ as the disparity filter does. Suppose $i$ has weighted directed edges connected to $k_i$ neighbors. Then, $p_{ij}$ is $w_{ij}/\sum_{m=1}^{k_i} w_{in_{im}}$ where $n_{im}$ is one of $i$'s neighbors. Similarly, $p_{ji}$ is $w_{ji}/\sum_{m=1}^{k_j} w_{jn_{jm}}$. $p_{ij} - p_{ji}$, the difference between local dependencies, is termed $D_{i \leftrightarrow j}$ that can range from -1 to 1. The value becomes close to 1 when $i$ fully depends on $j$ but a very weak dependence of $j$ on $i$ exists. $D_{i \leftrightarrow j}$ close to -1 is for the opposite case. $D_{i \leftrightarrow j} = 0$ indicates that relative local dependencies are same. To confirm whether $D_{i \leftrightarrow j}$ is statistically significant, the proposed method samples $x_{ij}$ from $Beta(1, k_i - 1)$ and $x_{ji}$ from $Beta(1, k_j - 1)$ and then calculates $x_{ij} - x_{ji}$ for a given number of trials. If $D_{i \leftrightarrow j}$ is substantially high or low compared to sampled $x_{ij} - x_{ji}$ values, the method adds a synthetic unweighted edge $i \to j$ or $j \to i$ to a backbone, respectively. Specifically, this operation happens when the proportion of sampled $x_{ij} - x_{ji}$ values above or below $D_{i \leftrightarrow j}$ is less than or equal to a certain threshold. A created edge from a node $s$ to a node $t$ in the backbone indicates that $t$ is at a higher position in a hierarchy than $s$. Note that this method cannot be applied when a node has only one neighbor (a Beta distribution requires parameters over 0) or two connected nodes have a unilateral relation. In this paper, values are sampled 10,000 times, and a threshold of 0.005 is set for the disparity-in-differences method (two-sided, leading to 0.01 in total to be matched with the threshold for the disparity filter).

    c. Empirical Weighted Directed Networks for Applications

The disparity-in-differences method was applied to a journal citation network, a U.S. airport network, the Enron email network, and a world trade network. Self-loops were removed in all networks. The journal citation network was constructed from an open bibliographic database, SciSciNet-v2 (Lin et al., 2023). Only the papers published in a journal of which name is in English and ISSN exists were included in the citation network. For each paper, its references were mapped to journals and then aggregated to weighted edges from the journal in which the paper was published to the journals in which its references were published. As the number of references varies by paper, edge weights were normalized by the sum of edge weights. These steps were taken for all papers, and resulting networks were aggregated into one large network. The journal citation network consists of 90,872 nodes and 76,512,317 edges. The U.S. airport network was constructed based on the 2024 Airline Origin and Destination Survey from the Bureau of Transportation Statistics of the U.S. Department of Transportation. Each edge weight of this network is the sum of passenger values from an origin airport to a destination airport, regardless of sequence numbers and coupons. The network contains 456 nodes and 17,287 edges. The Enron email network (Priebe et al., 2005) was downloaded directly from the igraphdata R package and processed to have email communication records made between the years 1998 and 2002. For the direct recipient type (i.e., "to"), the number of emails from an address to another address was counted and assigned as an edge weight. This network has 182 nodes and 2,829 edges. The world trade network was constructed by aggregating positive export values between the years 2020 and 2023 that were sourced from the international trade data (The Growth Lab at Harvard University, 2019). The network includes 235 nodes and 32,296 edges.

## 3. Results

Figure 1 shows how extracted backbones are different in size by varying thresholds for the empirical networks. Interestingly in all cases, the proposed method is less prone to threshold changes. When an extremely low threshold is used, the disparity-in-differences method still returns a backbone with a decent fraction of nodes, whereas the disparity filter returns a backbone with fewer nodes. In addition, extracted backbones by the disparity-in-differences method do not include all nodes and edges of the original networks even with a high threshold. It is because of unilateral relations or nodes with one neighbor in data so that statistical significance cannot be calculated for all node pairs by the disparity-in-differences method. Moreover, the proposed method converts bilateral relations between nodes to synthetic unilateral relations, resulting in fewer edges by design.

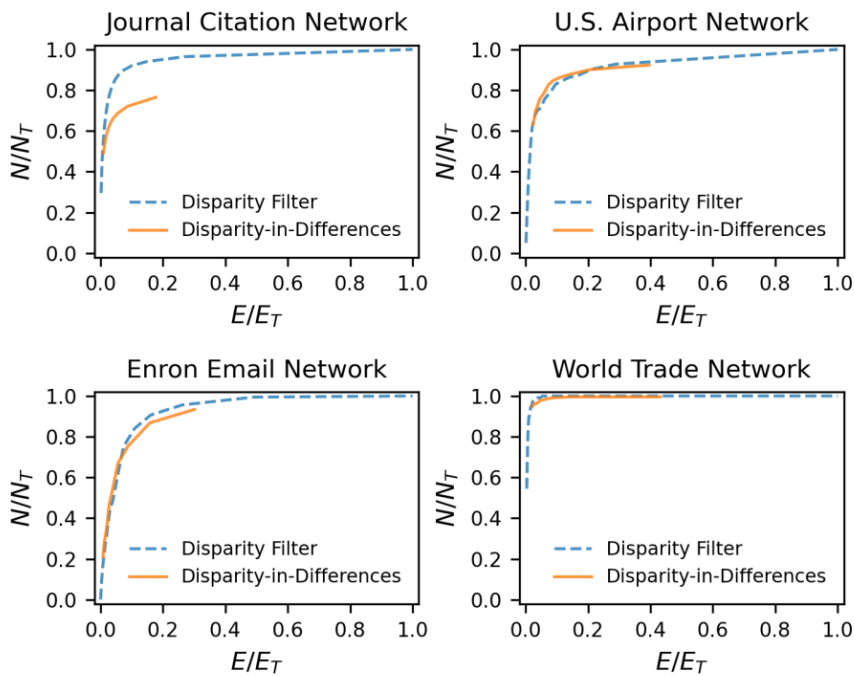

**Figure 1.** The fractions of nodes and edges in extracted backbones with varying thresholds compared to the original networks. $N_T$ and $E_T$ are the total numbers of nodes and edges in an original network, respectively.

To evaluate inferred relations, journal ratings (A*, A, B, C) by field of research (e.g., Information Systems) and airport hub size categories (Large, Medium, Small, Nonhub) were retrieved from the 2022 Australian Business Deans Council (ABDC) Journal Quality List and the National Plan of Integrated Airport Systems, respectively. If the disparity-in-differences method works as intended, its extracted backbones should be less likely to contain edges from higher-rank to lower-rank journals for each field of research and from larger to non- or smaller-size hub airports in a region. For this reason, the fraction of backbone edges from upper-level entities to lower-level entities (termed "misaligned proportion") was calculated as an evaluation metric for the backbones. The disparity-in-differences method returns lower misaligned proportion values than the disparity filter across all fields of research in business, except 3899 Other Economics (Figure 2 Top), and all regions in the U.S. airport network (Figure 2 Bottom). The high misaligned proportion of the disparity-in-differences method for the field of research 3899 would be due to less coherent bilateral relations between journals as the field name implies.

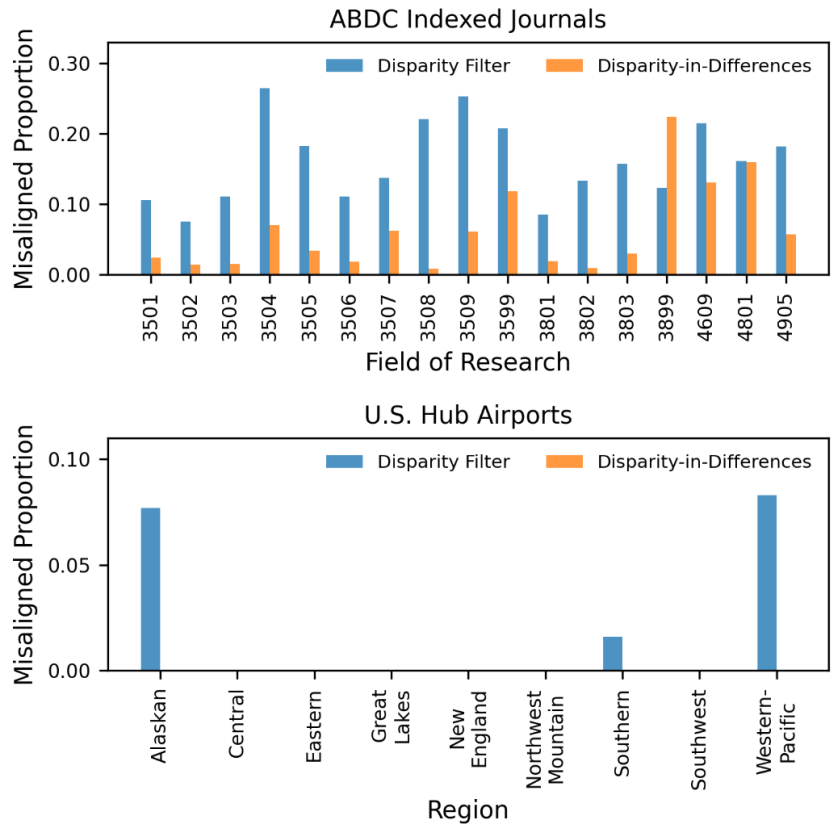

**Figure 2.** Misaligned proportions by fields of research in the ABDC Journal Quality List (top) and the regions in the U.S. airport network (bottom). No bar in a U.S. region means that all edges of a backbone correctly point from non- or smaller-size hub airports to larger-size hub airports in the region.

The proposed method also reveals a hierarchy between managers in the Enron email network though they are in different organizations and departments. Managers were grouped by role title into Executive (CEO, President), Senior (COO, Vice President, Director), and General (Manager) Management as classified in a previous study (Diesner et al., 2005). The backbone extracted by the disparity filter includes 3 misaligned edges from Executive to Senior Management and 1 misaligned edge from Senior to General Management, but no cases are found in the backbone extracted by the disparity-in-differences method.

Extracted backbones of the world trade network return similar lists of countries positioned high at a hierarchy with respect to the number of incoming edges. The U.S., China, and Germany are the top 3 countries in both backbones. A notable difference is the rank of India that is the 6th in the disparity filter backbone and the 4th in the disparity-in-differences backbone. The Netherlands ranks at the 4th in the disparity filter backbone. A case of Mexico shows that the disparity-in-differences method identifies hierarchical relations correctly. In the backbone built through the disparity filter, Mexico has one bilateral relation with the U.S., one outgoing edge to Canada, and eight incoming edges from Brazil, China, Columbia, Guatemala, Hong Kong, Honduras, Nicaragua, and Trinidad and Tobago. The disparity-in-differences method infers that the U.S. and Canada are at a higher position and Columbia, Guatemala, Honduras, and Nicaragua are at a lower position than Mexico. These relations are consistent with a core-periphery structure of countries (Chase-Dunn et al., 2000). The proposed method determines that Mexico's connections with Brazil, China, Hong Kong, and Trinidad and Tobago are not statistically asymmetrical.

## 4. Discussion

The disparity-in-differences method extends the disparity filter to uncover hierarchical relations with strong theoretical and methodological backgrounds. Its applications to four empirical networks validate the method and suggest possibilities to be widely used in various disciplines. For example, the proposed method may generate hierarchical backbones in biological systems (Pathak et al., 2024). A big advantage of the proposed method is to supplement the disparity filter not to replace it. Statistically significant edges extracted from the disparity filter should be considered when analyzing complex systems as the current literature has done, while their asymmetrical dependencies extracted from the disparity-in-differences method would offer additional perspectives.

Although used for evaluation, misaligned edges can provide further implications. First, misaligned edges would be associated with external factors not captured in weighted directed networks but explained by domain knowledge. Exploring these factors will enhance our understanding of complex systems. Second, misaligned edges would be observed when used hierarchical categories do not fully represent the currently perceived hierarchy. A backbone extracted from the disparity-in-differences method can address this issue by characterizing unseen or emerging nodes based on their neighbors in the given hierarchy. Graph embedding on a hyperbolic space seems to be a good option for the task as it encodes both hierarchy and similarity of nodes (Nickel & Kiela, 2017). This type of techniques would be more feasible once closed loops that the disparity-in-differences method may generate are properly handled to have a backbone in a directed acyclic graph format.

**Data, Documentation, and Code Availability**

All materials for this manuscript are available at https://github.com/hkim07/Disparity-in-Differences.

**References**

Casciaro, T., & Piskorski, M. J. (2005). Power imbalance, mutual dependence, and constraint absorption: A closer look at resource dependence theory. *Administrative Science Quarterly*, 50(2), 167-199.

Chase-Dunn, C., Kawano, Y., & Brewer, B. D. (2000). Trade globalization since 1795: Waves of integration in the world-system. *American Sociological Review*, *65*(1), 77-95.

Coscia, M., & Neffke, F. M. (2017, April). Network backboning with noisy data. In *2017 IEEE 33rd International Conference on Data Engineering (ICDE)* (pp. 425-436). IEEE.

Dianati, N. (2016). Unwinding the hairball graph: Pruning algorithms for weighted complex networks. *Physical Review E*, *93*(1), 012304.

Diesner, J., Frantz, T. L., & Carley, K. M. (2005). Communication networks from the Enron email corpus "It's always about the people. Enron is no different". *Computational & Mathematical Organization Theory*, *11*(3), 201-228.

Emerson, R. M. (1962). Power-dependence relations. *American Sociological Review*, 27(1), 31-41.

The Growth Lab at Harvard University. (2019). *International Trade Data (HS, 92)* (Version V15) [dataset]. Harvard Dataverse. https://doi.org/10.7910/DVN/T4CHWJ

Lin, Z., Yin, Y., Liu, L., & Wang, D. (2023). SciSciNet: A large-scale open data lake for the science of science research. *Scientific Data*, 10(1), 315.

Marcaccioli, R., & Livan, G. (2019). A Pólya urn approach to information filtering in complex networks. *Nature Communications*, *10*(1), 745.

Nickel, M., & Kiela, D. (2017). Poincaré embeddings for learning hierarchical representations. *Advances in Neural Information Processing Systems*, *30*.

Pathak, A., Menon, S. N., & Sinha, S. (2024). A hierarchy index for networks in the brain reveals a complex entangled organizational structure. *Proceedings of the National Academy of Sciences*, *121*(27), e2314291121.

Priebe, C. E., Conroy, J. M., Marchette, D. J., & Park, Y. (2005). Scan statistics on Enron graphs. *Computational & Mathematical Organization Theory*, *11*(3), 229-247.

Serrano, M. Á., Boguná, M., & Vespignani, A. (2009). Extracting the multiscale backbone of complex weighted networks. *Proceedings of the National Academy of Sciences*, 106(16), 6483-6488.